\documentclass{aa}
\usepackage{hyperref}
\usepackage{txfonts}
\usepackage{graphicx}
\usepackage[tbtags,fleqn]{amsmath} 
\usepackage{multirow}
\usepackage{float}
\usepackage{fancyhdr}

\bibpunct{(}{)}{;}{a}{}{,} 

\def\1702{4U 1702-429}
\def\xmm{{\it XMM-Newton }}
\def\int{{\it INTEGRAL }}

\begin{document}


\title{Study of the reflection spectrum of the LMXB 4U 1702-429 }

\author{R. Iaria\inst{1}, T. Di Salvo\inst{1}, M. Del Santo\inst{2},
  F. Pintore\inst{3}, A. Sanna\inst{4}, A. Papitto\inst{5},
  L. Burderi\inst{4}, A. Riggio\inst{4}, A. F. Gambino\inst{1}, M. Matranga\inst{1}}

\institute{Dipartimento di Fisica e Chimica,
 Universit\`a di Palermo, via Archirafi 36 - 90123 Palermo, Italy\\
\email{rosario.iaria@unipa.it}
\and 
Istituto Nazionale di Astrofisica, IASF Palermo, Via U. La Malfa 153,
I-90146 Palermo, Italy
\and
INAF-Istituto di Astrofisica Spaziale e Fisica Cosmica - Milano, via E. Bassini 15, I-20133 Milano, Italy 
\and
Dipartimento di Fisica, Universit\`a degli Studi di Cagliari, SP
Monserrato-Sestu, KM 0.7, Monserrato, 09042 Italy
\and
Institute of Space Sciences (ICE, CSIC-IEEC), Carrer de Can Magrans, S/N, E-08193 Barcelona, Spain}

\date{\today}

\abstract
{The source \object{\1702} (Ara X-1) is a low-mass X-ray binary system
  hosting a neutron star. Albeit the source is quite bright ($\sim10^ {37}$ erg s$^{-1}$) its  broadband
  spectrum has  never been studied. Neither dips nor eclipses have
  been observed in the light curve suggesting  that its inclination angle is
  smaller than 60$^{\circ}$. } 
{We analysed the broadband spectrum of \1702 in the 0.3-60 keV energy range, using {\it XMM-Newton}
  and {\it INTEGRAL} data,  to constrain its Compton reflection component if
  it is present. }
{After excluding the three time intervals in which three type-I X-ray
    bursts occurred, we  fitted the joint XMM-Newton and INTEGRAL 
spectra obtained from   simultaneous observations.}
{ A broad emission line at 6.7 keV and  two absorption edges at 0.87 and 8.82
keV were detected.  We found that a self-consistent reflection model
fits the 0.3-60 keV spectrum well. The broadband continuum is composed of
 an emission component originating  from the inner region of the
 accretion disc,  a Comptonised  direct emission coming from a corona
 with an electron temperature of  $2.63 \pm 0.06$ keV and an optical
 depth $\tau=13.6 \pm 0.2$, and, finally, a reflection component. 
The best-fit indicates that the broad emission line and 
the absorption edge at 8.82 keV, both associated with the presence of
\ion{Fe}{xxv} ions, are produced by reflection in the region above  the
disc with a  ionisation parameter of Log$(\xi) \simeq 2.7$. 
We have inferred that the inner  radius, where the broad emission line  
originates,  is $64^{+52}_{-15}$ km, and the inner radius of the
accretion disc is $39^{+6}_{-8}$ km. The emissivity of the reflection
  component and the inclination angle of the system are  
    $r^{-3.2^{+0.5}_{-5.1}}$ and  $44^{+33}_{-6}$ degrees,
  respectively. The absorption edge at 0.87 keV is associated to the
  presence of \ion{O}{viii} ions and it is
  produced in a region above the disc with Log$(\xi) \simeq 1.9$}. 
{}
  \authorrunning{R.\ Iaria et al.}

\titlerunning{The broadband spectrum analysis of 4U 1702-429}

\maketitle

\section{Introduction}
Low-mass X-ray binaries (LMXBs) usually show discrete features such as emission lines and absorption
edges. The most prominent feature is an emission line at 6.4-6.97 keV,
usually interpreted as a fluorescence line from iron at different ionisation
states. In fact, iron is a relatively abundant element with the highest
fluorescence yield among the most abundant atomic species. These features
are powerful tools to investigate the structure of the accretion flow
close to the central source; in particular, important information can
be obtained from the detailed spectroscopy of the line profile, since
it is determined by the ionisation state, geometry, and velocity field
of the reprocessing plasma \citep[see][for a review]{cackett_2010}.
 
These emission lines are usually broad with Gaussian
$\sigma$ from 0.3 up to more than 1 keV. This broadness is
incompatible with a simple thermal broadening caused by the plasma
temperature, because of the large mass of iron atoms. It has been
interpreted as being caused by Compton broadening in a Comptonising medium
of moderate temperatures and optical depth \citep[see e.g.][]{ng_10}
or Compton scattering caused by strong outflowing winds illuminated by
the radiation from the innermost part of the system \citep{tita_09}.
Similarities were found between the accretion flows and the overall
spectral shapes in LMXBs hosting neutron stars (NSs) and black holes (BHs). In both systems, an
accretion disc surrounds a Comptonising corona located around the
compact objects. This has led to the conclusion that in both LMXB
systems, these emission lines may be produced by reflection of the
primary continuum over the inner accretion disc.  In this scenario, the
line profile is shaped by Doppler and relativistic effects caused by
the fast (Keplerian) motion of the plasma in the inner regions of the
accretion disc.  As a consequence, the line shows a characteristic
broad and asymmetric (red-skewed) profile, the detailed shape of which
depends on the inclination of the system with respect to the line of
sight, and on how deep the accretion disc extends into the NS
gravitational potential \citep[see][]{fabian_89,matt_92}.

If the origin of this line is from disc reprocessing, one would also
expect the presence in the spectra of other discrete features (such
as emission lines and absorption edges from the other abundant elements)
and an excess of emission (Compton hump) caused by direct Compton
scattering of the primary spectrum by the electrons in the disc.
Indeed, broad emission lines from Silicon, Argon, and Calcium have been detected
together with iron features in the spectra of bright NS LMXBs (such as
4U 1705-44, e.g. \citealt{disalvo_09}; GX 349+2, \citealt{iaria_09};
GX340+0, \citealt{dai_09}; GX 3+1, \citealt{piraino} and
\citealt{pintore_15}), and
in some cases, a broadened absorption edge at $8-8.5$ keV was also
required. The ionisation states of these elements were compatible
with similar values of the ionisation parameter $\xi$, and
the ratios of the widths of these features with respect to
the corresponding energy were compatible with being constant for each
source, implying that all these features were 
produced in the same disc region.
The Compton hump at 20-40 keV has also been detected in the hard
spectral state of these sources with high statistical significance
\citep[see e.g.][]{disalvo_15,miller_13,degenaar_15,piraino_16}, in combination
with the presence of the iron line, and both these features have been
modelled with  self-consistent reflection models.
The reflection model is able to simultaneously fit all these features
(broad emission lines and absorption edges as well as the Compton hump)
and is therefore the most promising explanation for their origin 
\citep[see e.g.][]{disalvo_15,dai_10,reis_09,cackett_2010}. 

The X-ray source 4U 1702-429 (Ara X-1) is a NS LMXB showing type-I
X-ray bursts. The source was detected as a burster
with \textit{OSO 8} \citep{swank_76}, whilst the persistent
X-ray emission was detected by \cite{lewin_79}. \cite{Oo_91}
classified 4U 1702-429 as an atoll source using EXOSAT data. 
Using Chandra HRC-I  data, \cite{wachter_05} gave the accurate position
of the X-ray source  with an associated error of $0\arcsec\!.6$.  \cite{galloway_08},
analysing  the photospheric radius expansion
during the observed type-I X-ray bursts,   inferred a
distance to the  source of $4.19 \pm 0.15$ kpc and   $5.46 \pm 0.19$ kpc
for a pure hydrogen and pure helium companion star, respectively. 
 Furthermore, these authors suggested that the companion star
 should have a  mass fraction of hydrogen lower than 50\%. 
 \cite{mark_99}, using the data of the proportional counter array
 (PCA) onboard  the {\it Rossi-XTE} (RXTE) satellite, detected burst
oscillations at 330 Hz that could be associated with the spin
frequency of the NS.  

Up to now a few works reported the  analysis of the persistent
spectrum of 4U 1702-429. \cite{Cristian_97} analysed three
observations taken by the {\it Einstein} satellite combining the data
of the solid-state spectrometer (SSS; 0.5-4.5 keV) and of the monitor
proportional counter (MPC; 1.2-20 keV).  The authors fitted the
spectrum of the source with an absorbed cut-off power-law obtaining an equivalent
hydrogen column density $N_H$ of the interstellar medium between
$1.1 \times 10^{22}$ and $1.7 \times 10^{22}$ cm$^{-2}$. The
photon-index spanned the range between 1.3 and 1.5, and the cut-off
temperature between 8 and 16 keV.
 \cite{mark_99} analysed three observations of 4U 1702-429 
taken with  RXTE/PCA. These authors fitted the persistent spectrum with
a cut-off power-law inferring a cut-off-temperature between 3.5 and 4.6
keV. 
\begin{figure*}
\centering
\includegraphics[width=8.5cm]{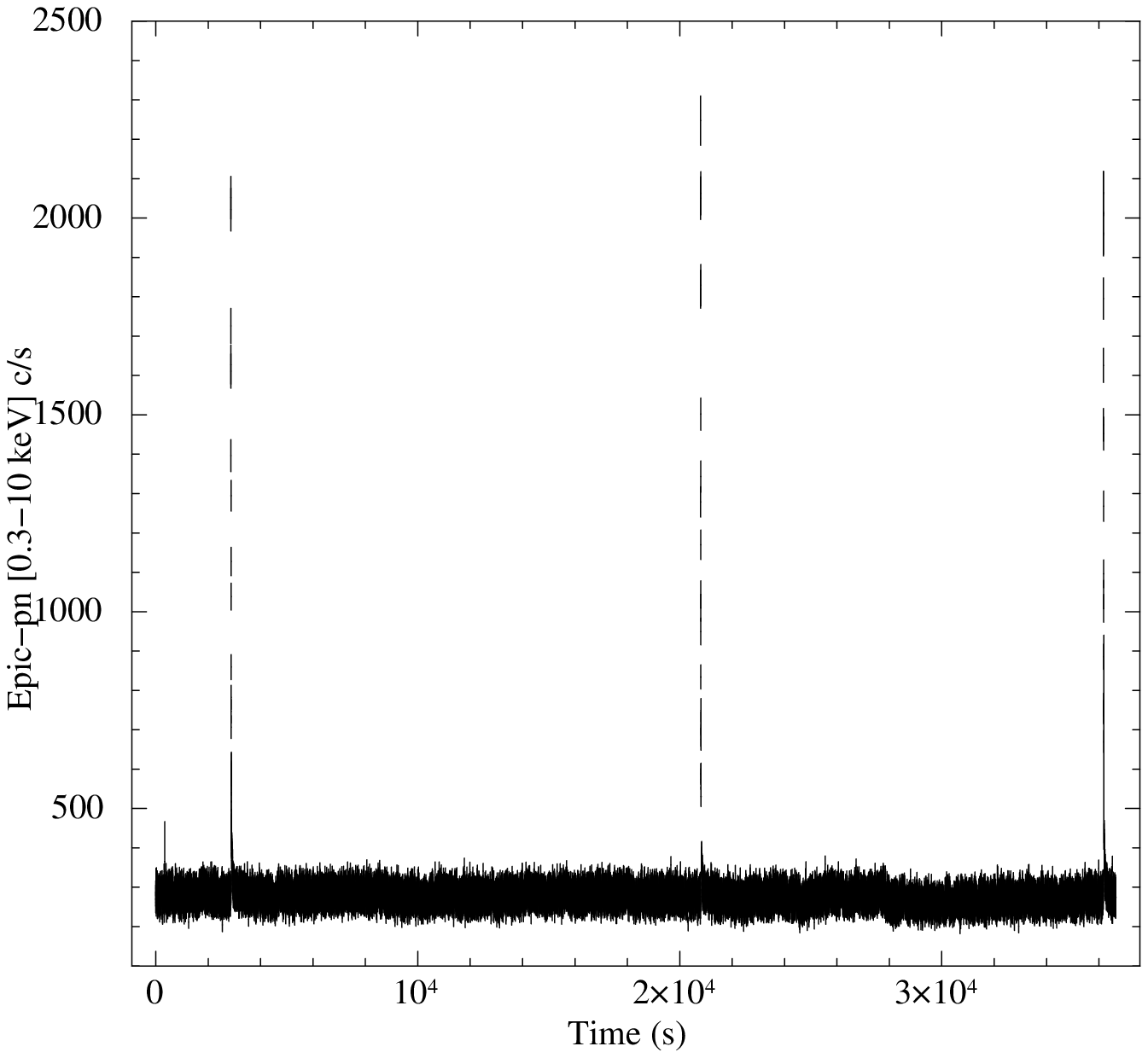}\hspace{0.3truecm}
\includegraphics[width=8.5cm]{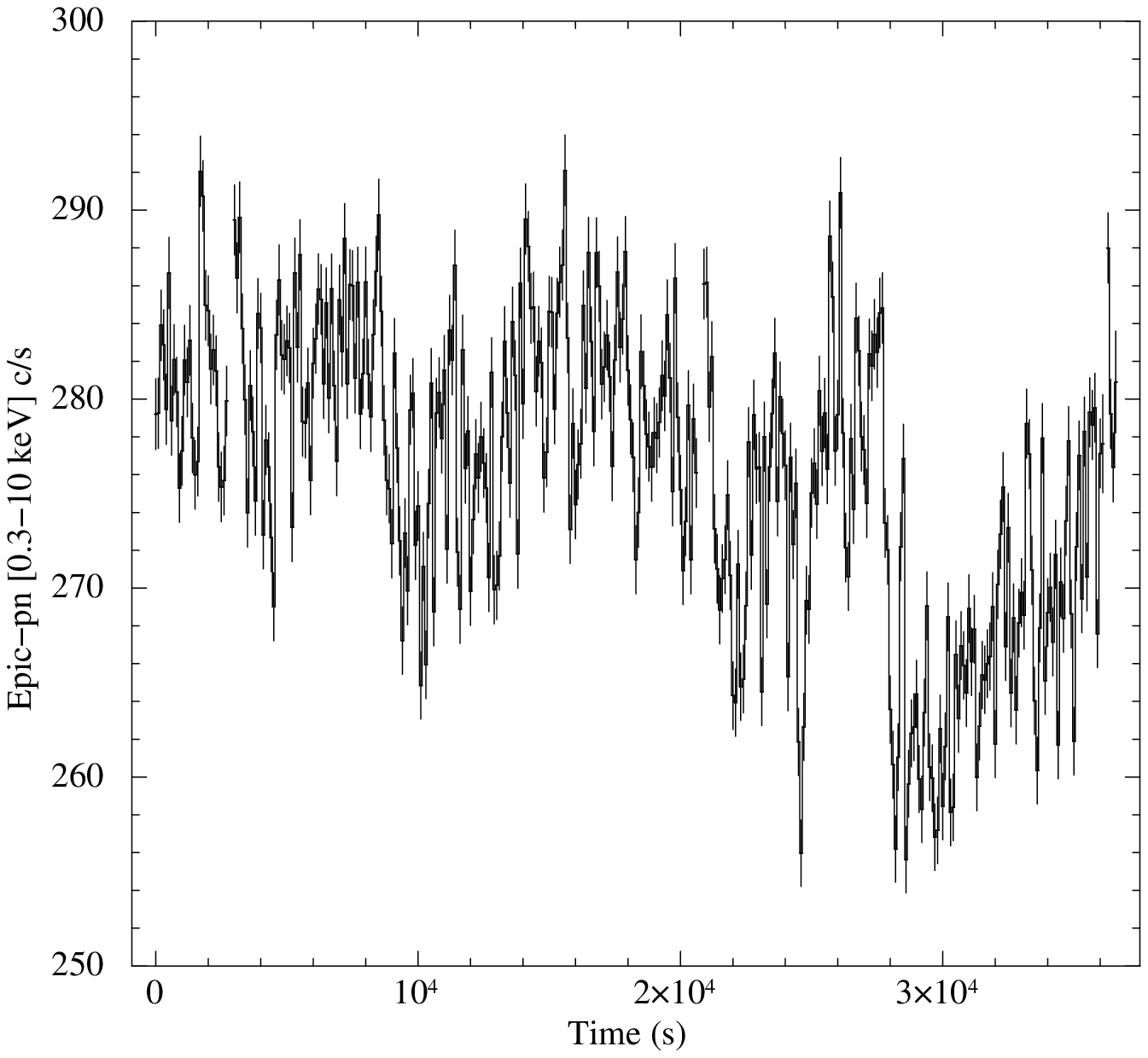}\\
\includegraphics[width=8.5cm]{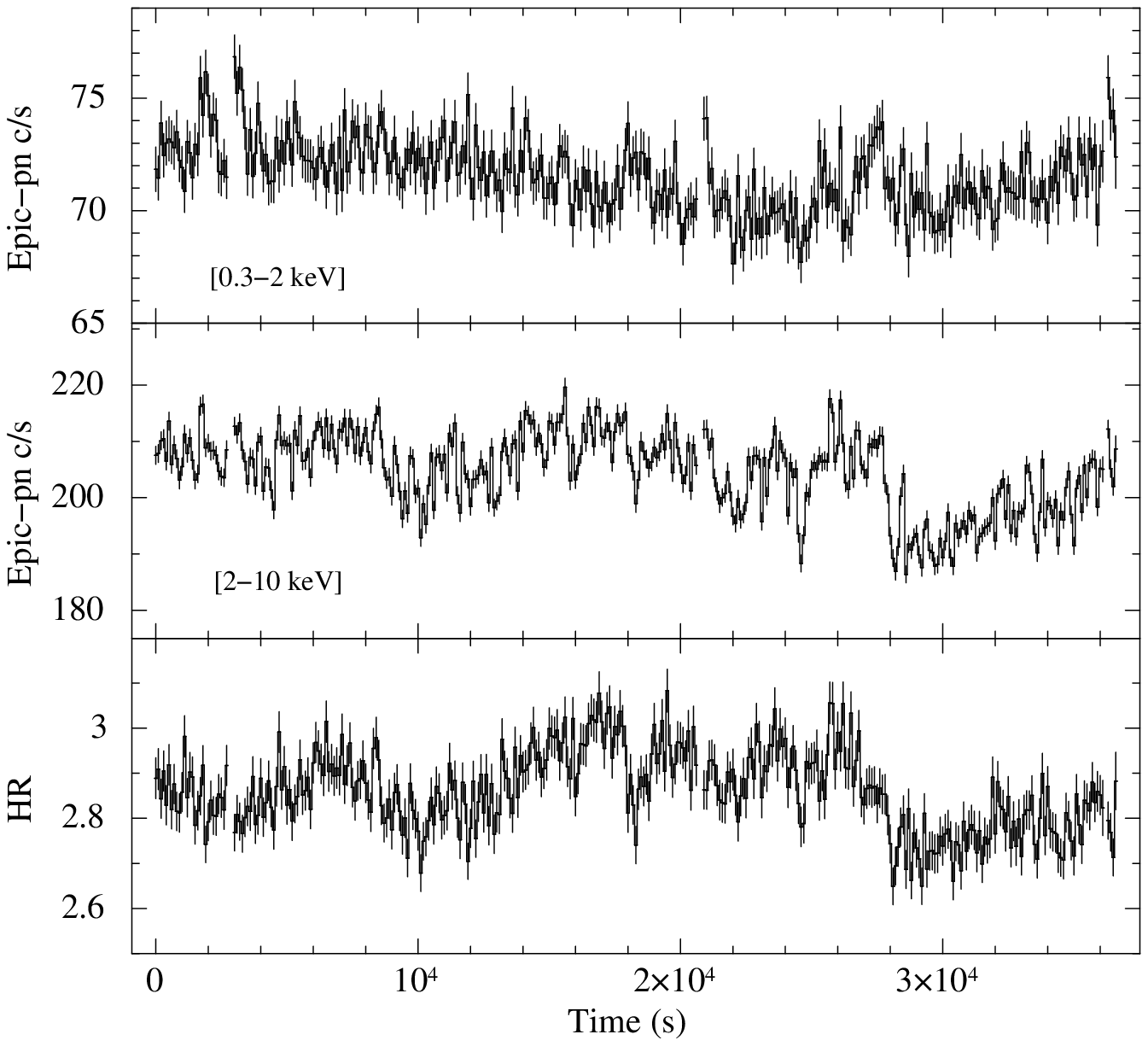}\hspace{0.3truecm}
\includegraphics[width=8.5cm]{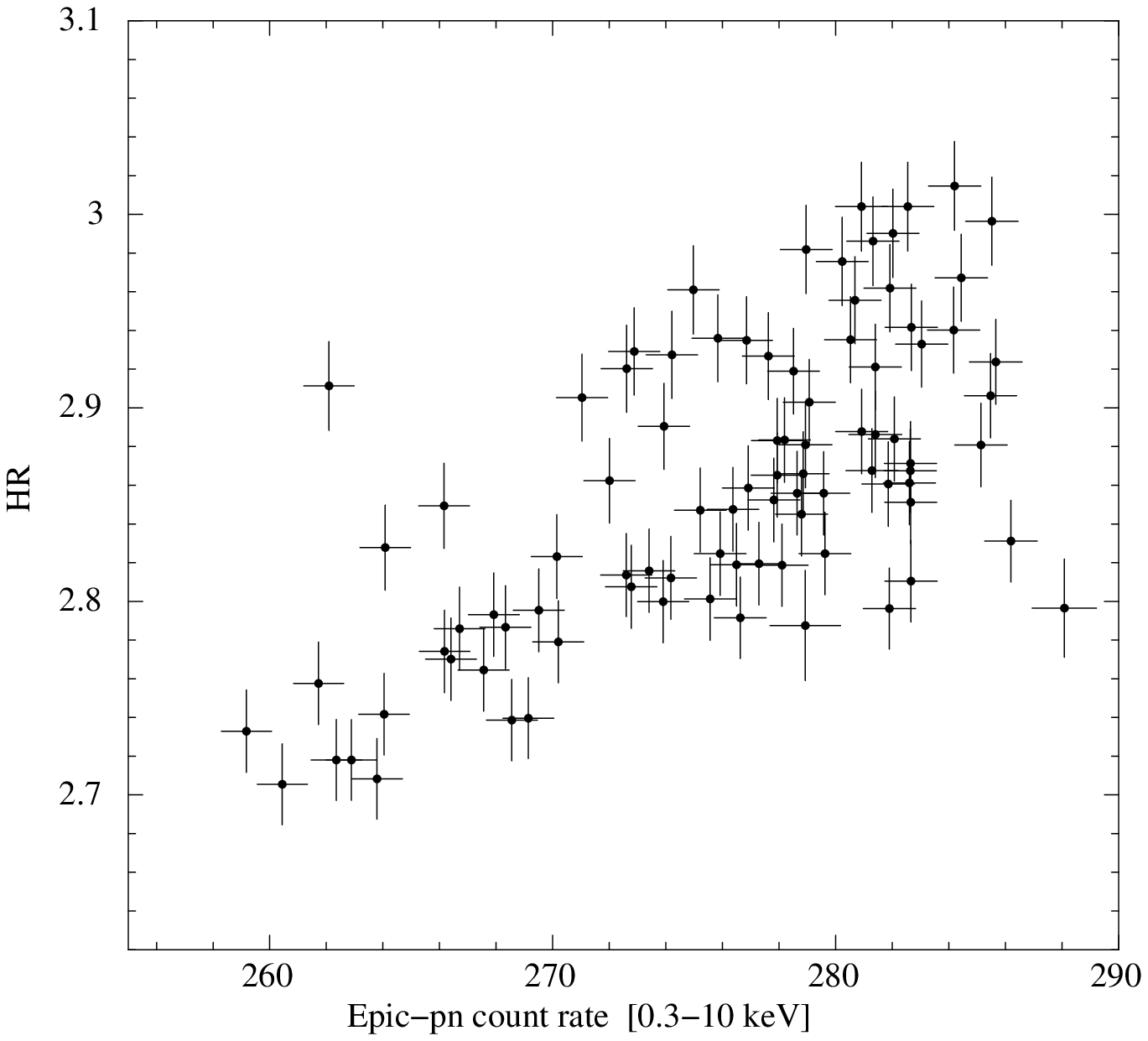}
\caption{Top-left panel: EPIC-pn light curve of \1702 with a bin time of 1
  s. Three type-I X-ray bursts occur  during the observation: between
  2860 and    2980 s, 20780 and  20900 s, and
  36150 and   36300 s from the start time.
Top-right panel: 0.3-10 keV    EPIC-pn persistent light curve.   The bin time is
100 s.
Bottom-left panel: 0.3-2 keV EPIC-pn persistent light curve (upper panel),  2-10
  keV EPIC-pn persistent light curve (middle panel) and the
  corresponding HR (lower panel). The bin time is
100 s.  Bottom-right panel: HR vs. intensity (bin time of 400 s). }
 \label{pn_light}
\end{figure*}
In this work we show the spectral analysis of the persistent spectrum
of 4U 1702-429  in the 0.3-60 keV energy range  
using both  {\it XMM-Newton}  and  {\it INTEGRAL}  observations.

\section{Observations and data reduction}

The \xmm observatory \citep{Jansen_01} includes three 1500
cm$^2$ X-ray telescopes each with an European Photon Imaging Camera
(EPIC, 0.1-15 keV) at the focus. Two of the EPIC imaging spectrometers
use MOS charge-coupled devices CCDs \citep{turner_01} and one uses pn CCDs
\citep{struder_01}.  Reflection grating spectrometers \citep[RGS,
0.35-2.5 keV,][]{herder_01} are located behind two of the
telescopes. 

The region of sky containing \1702 was observed by \xmm between 2010
March 09 14:11:26 UT and March 10 00:22:08 UT (OBSid 0604030101) for a
duration of 38 ks. During the observation, the MOS1 and MOS2
instruments were operating in Small window mode, while the EPIC-pn camera was
operating in Timing mode with medium filter. 
To reduce the \xmm data, we used the Science Analysis Software (SAS) v. 14.0.0 and the
calibration files at the date of 2014 April 25.  Initially we
produced the EPIC-pn light curve in the 0.3-10 keV energy range.  We
extracted the 0.3-10 keV EPIC-pn events of the source selecting RAWX
between 24 and 48 and setting PATTERN$\le$4 (single and double pixel
events) and FLAG=0. The 0.3-10 keV background events were extracted by
selecting RAWX between two and 12.  Then, we used the {\tt epiclccorr} to
obtain the background-subtracted EPIC-pn light curve that is shown in
Fig. \ref{pn_light} (top-left panel) with a bin time of 1 s.  Three
type-I X-ray bursts occurred during the observation, between 2860 and
2980 s, 20780 and 20900 s, and 36150 and 36300 s from the start time.
The persistent emission has a count rate of 280 c s$^{-1}$ whilst during the
bursts the count rate increases to 2100 c s$^{-1}$.  Since the aim of this
work is the spectral analysis of the persistent emission, we excluded
the time intervals containing the bursts in all the analysed datasets. The light curve of the
persistent emission is shown in Fig.  \ref{pn_light} (top-right
panel) and the bin time is 100 s. The light curve shows a count rate
almost constant at 280 c s$^{-1}$ up to 28000 s from the start time, then the
count rate suddenly decreases at 260 c s$^{-1}$ and gradually increases again
up to the end of the observation, coming back to 280 c s$^{-1}$. Since
 the maximum count rate to avoid  pile-up issues is  800 c s$^{-1}$ for
the EPIC-pn camera used in Timing mode, we are confident that the
 persistent EPIC-pn spectrum is not affected by pile-up. 
  Using the SAS tool {\tt epatplot,}  we found
   that the pile-up fraction for single and double events is only 0.1\% 
and 3.3\%, respectively, in the 2.4-10 keV energy range.
  
We also extracted the 0.3-2 keV and 2-10 keV EPIC-pn light curves and
produced the corresponding hardness ratio (HR, see bottom-left panel
in Fig. \ref{pn_light}). The HR value is quite constant at 2.85 from
the beginning of the observation up to 28 ks; after this, its value ranges
between 2.65 and 2.8, suggesting that the spectral shape is softer in
the last 10 ks of the observation.  Finally, we show the HR vs. the
0.3-10 keV EPIC-pn count rate in Fig. \ref{pn_light} (bottom-right
panel); the HR increases when the count rate increases.
  
\begin{figure*}
\centering
\includegraphics[width=8.5cm]{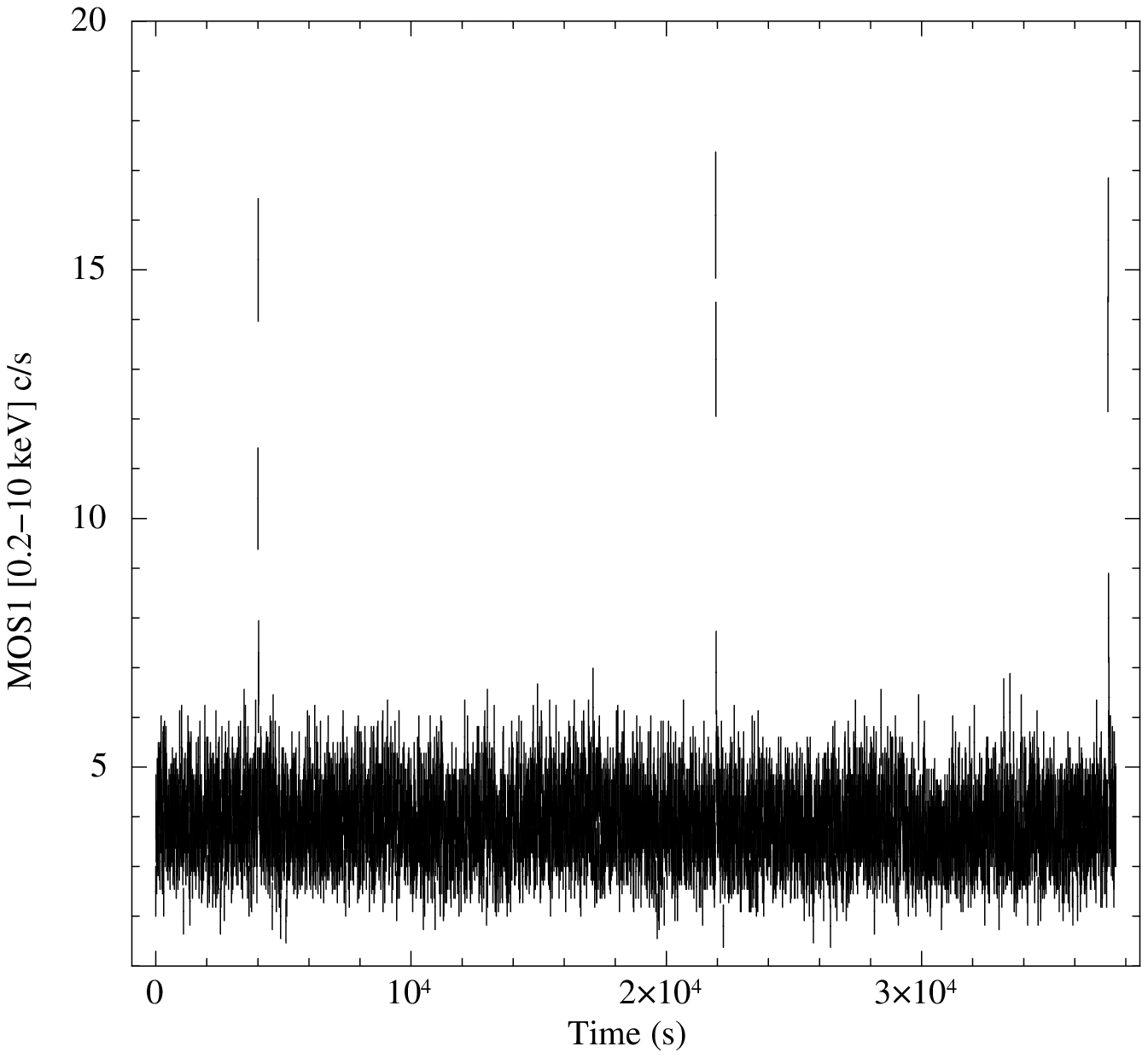}\hspace{0.3truecm}
\includegraphics[width=8.5cm]{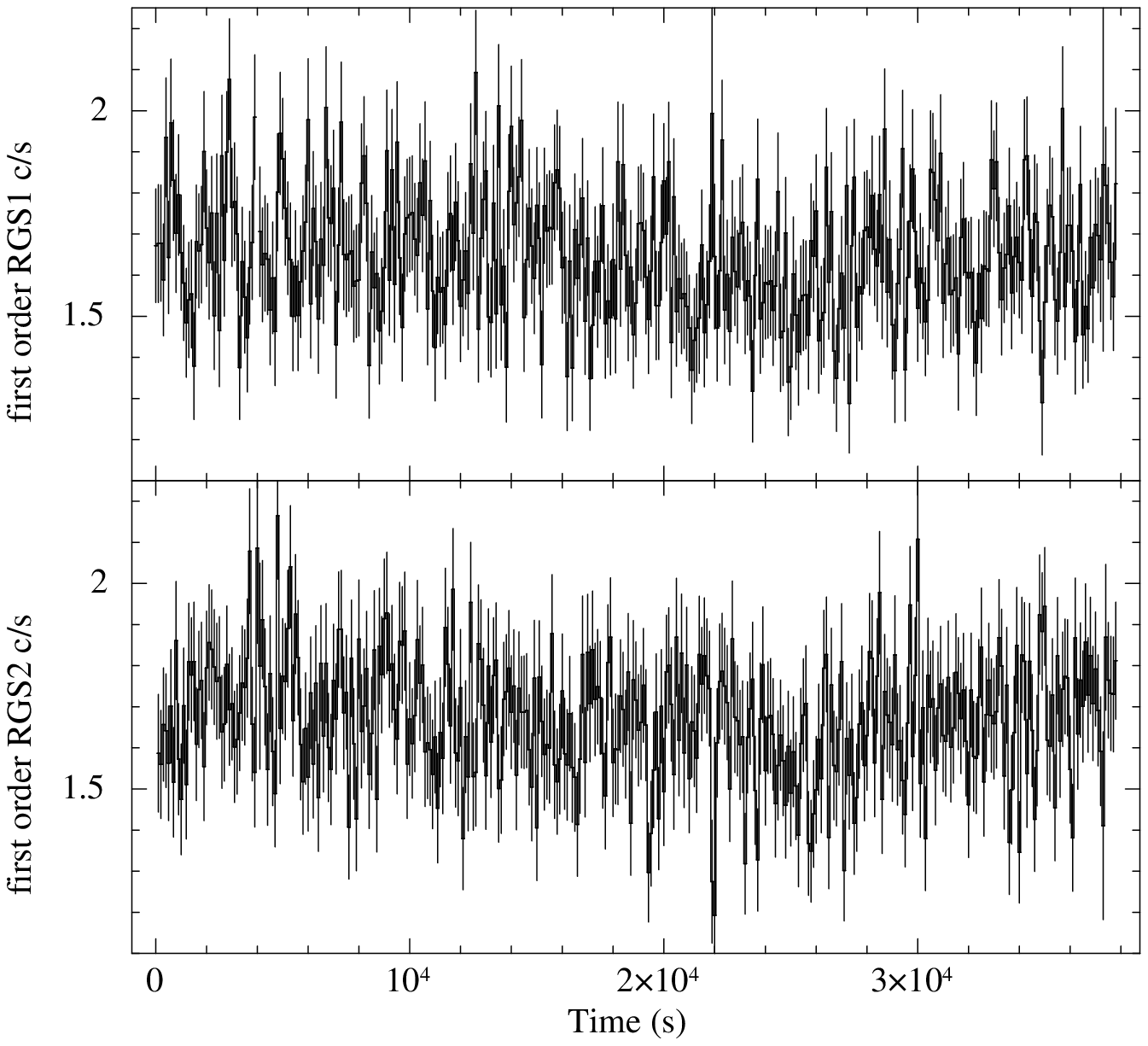}\\
\caption{Left panel: 0.2-10 keV MOS1 light curve obtained extracting the events
  from an annular region with inner and outer radius of 35$\arcsec\!$ and
  45$\arcsec\!$. The bin time is 10 s. Right panel: the first-order RGS1
  and  RGS2 light curves (bursts excluded) with a bin time
  of 100 s.}
 \label{mos_lc}
\end{figure*}

The MOS1 and MOS2 light curves were initially extracted using a
circular region centred to the source position and with a radius of
45$\arcsec$. The count rate of the 0.2-10 keV MOS1 and MOS2 light
curves is close to 50 c s$^{-1}$. Since the maximum count rate to
avoid pile-up issue is 4.5 c s$^{-1}$ for MOS cameras used in Small window
mode\footnote{http://xmm2.esac.esa.int/docs/documents/CAL-TN-0200-1-0.pdf.}, 
we expected  the presence of heavy pile-up in this case. Using
  the SAS tool
  {\tt epatplot,}  we estimated a large pile-up fraction of 20\%  in
  the 0.2-10 keV energy range.
To minimise the pile-up issues, we extracted the MOS1 and MOS2  
 source events from an annular region centred on the X-ray position of  the
 source and
with an inner and outer radius of 35$\arcsec$ and 45$\arcsec\!$,
respectively. In this case the persistent count rate
is 3.7 c s$^{-1}$ and the pile-up fraction is less than 2\% for MOS1 and MOS2 events in the
0.2-10 keV energy range. We show the 0.2-10 keV MOS1 light curve in
Fig. \ref{mos_lc} (left panel).

Finally, we  extracted the first order  RGS1 and RGS2 light curves
of the persistent emission. The count rate of the  RGS1 and RGS2 light
curves is 1.63 and 1.66 c s$^{-1}$, respectively, during the persistent
emission. We show the two light curves excluding the bursts in
Fig. \ref{mos_lc} (right panel).

To analyse the persistent spectrum of \1702, we extracted the EPIC-pn
spectrum from events with PATTERN$\le$4 and FLAG=0.  Source and
background spectra were extracted selecting the ranges
RAWX=[24:48] and RAWX=[2:12], respectively. EPIC-pn spectra were
rebinned in energy with an oversample of three using the SAS tool {\tt
  specgroup}.  MOS1 and MOS2 spectra were extracted from events with
PATTERN$\le$12 and FLAG=0. 
 The MOS1 and
MOS2 background spectra were obtained using the same settings and the
same annulus centred in a source-free region.  Furthermore, using the
SAS tool {\tt epatplot}, we found that the fraction of pile-up is close
to 3\% in MOS1 and MOS2 spectra. After checking that the MOS1
and MOS2 spectra were similar and that systematic features were absent
in the 0.3-10 keV energy range, we  used the SAS tool {\tt
  epicspeccombine} to obtain the merged spectrum (hereafter MOS12
spectrum). The MOS12 spectrum was grouped with a minimum of 25
counts per energy channel.  

We extracted the first-order RGS spectra
using the SAS tool {\tt rgsproc} and excluding the time intervals in
which the three bursts occur. After  checking that the RGS1 and
RGS2 spectra were similar and without systematic features in the
0.5-2 keV energy range, we  used the SAS tool {\tt rgscombine} to
obtain the merged spectrum (hereafter RGS12 spectrum).  The RGS12
spectrum was grouped with a minimum of 25 counts per energy
channel.
The effective times of the persistent
spectra are 37 ks for the RGS12 spectrum and 36 ks for MOS12
and EPIC-pn spectra, respectively.

We  searched for INTEGRAL observations performed on 2010 March,
both with the JEM-X \citep{lund03} and IBIS \citep{ubertini03}.  We selected three JEM-X pointings, namely science
windows (SCW), performed in March 10 and 11, and 58 IBIS SCWs
performed in March between 5 and 15.  The data-set selection
criterion was based on the maximisation of the spectral response of the
two telescopes, obtained with the source located within 4.5$^{\circ}$
and 3.5$^{\circ}$ from the centre of the IBIS and JEM-X FOVs,
respectively.  Thus, we  analysed IBIS/ISGRI \citep{lebrun03} and
JEM-X2 (JEM-X1 was switched-off at that time) data with the INTEGRAL
standard software OSA 10.1 \citep{courvoisier03}.

A JEM-X2 spectrum was extracted in 16 channels and an effective
exposure time of 6 ks was derived.  Because of the faintness of
the source in hard X-rays, the IBIS/ISGRI spectrum was extracted
by the mosaic of the pointing images.  We obtained a total mosaic in three
energy ranges, 20-26 keV, 26-37 keV, and 37-60 keV, and an
IBIS/ISGRI spectrum (130 ks effective exposure) by using the {\tt
  mosaic\_spec} tool on this image. We note that, based on the latest calibration issues, 
the JEM-X2 spectrum was used from 5 keV up to 22 keV.

We fitted simultaneously the \xmm (RGS12,  MOS12, and EPIC-pn)
and \int (JEM-X2 and ISGRI) spectra  using XSPEC v. 12.8.2. 
We selected the energy ranges 0.6-2.0 keV for RGS12, 0.3-10
keV for MOS12, 2.4-10 keV for  EPIC-pn, 5-25 keV  for JEM-X2, and  20-50 keV for ISGRI. 
We ignored the EPIC-pn energies channels lower than
2.4 keV since   a soft excess  is present that is not reconciled with
the RGS12 and MOS12 spectra, suggesting that the EPIC-pn calibrations are still 
uncertain below this energy (internal XMM-Newton report CAL-TN- 2
0083\footnote{http://xmm2.esac.esa.int/docs/documents/CAL-TN-0083.pdf.};
see also \citealt{piraino, pintore_15}).
\begin{figure*}
\centering
\includegraphics[width=8.5cm]{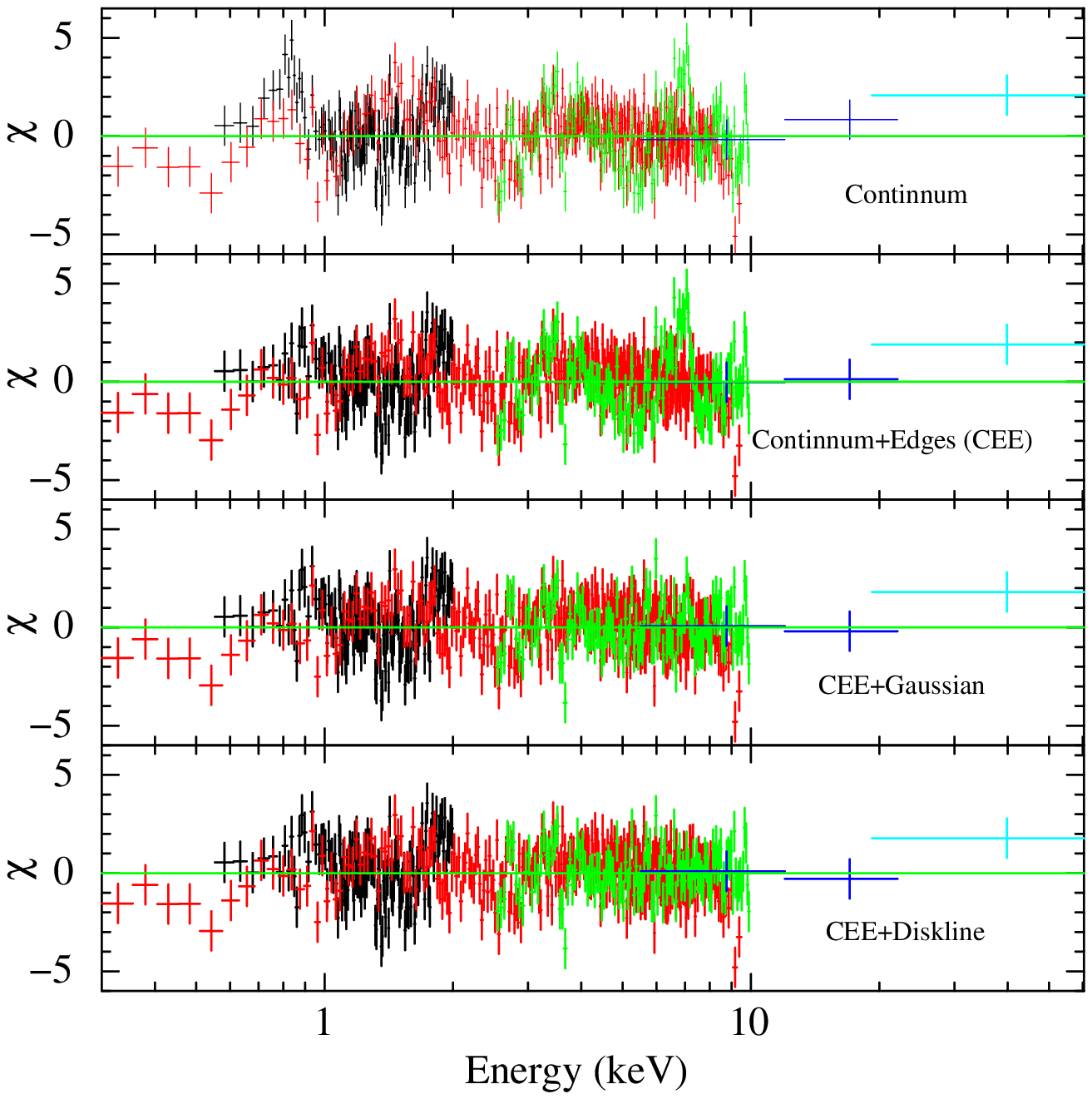}\hspace{0.3truecm}
\includegraphics[width=8.5cm]{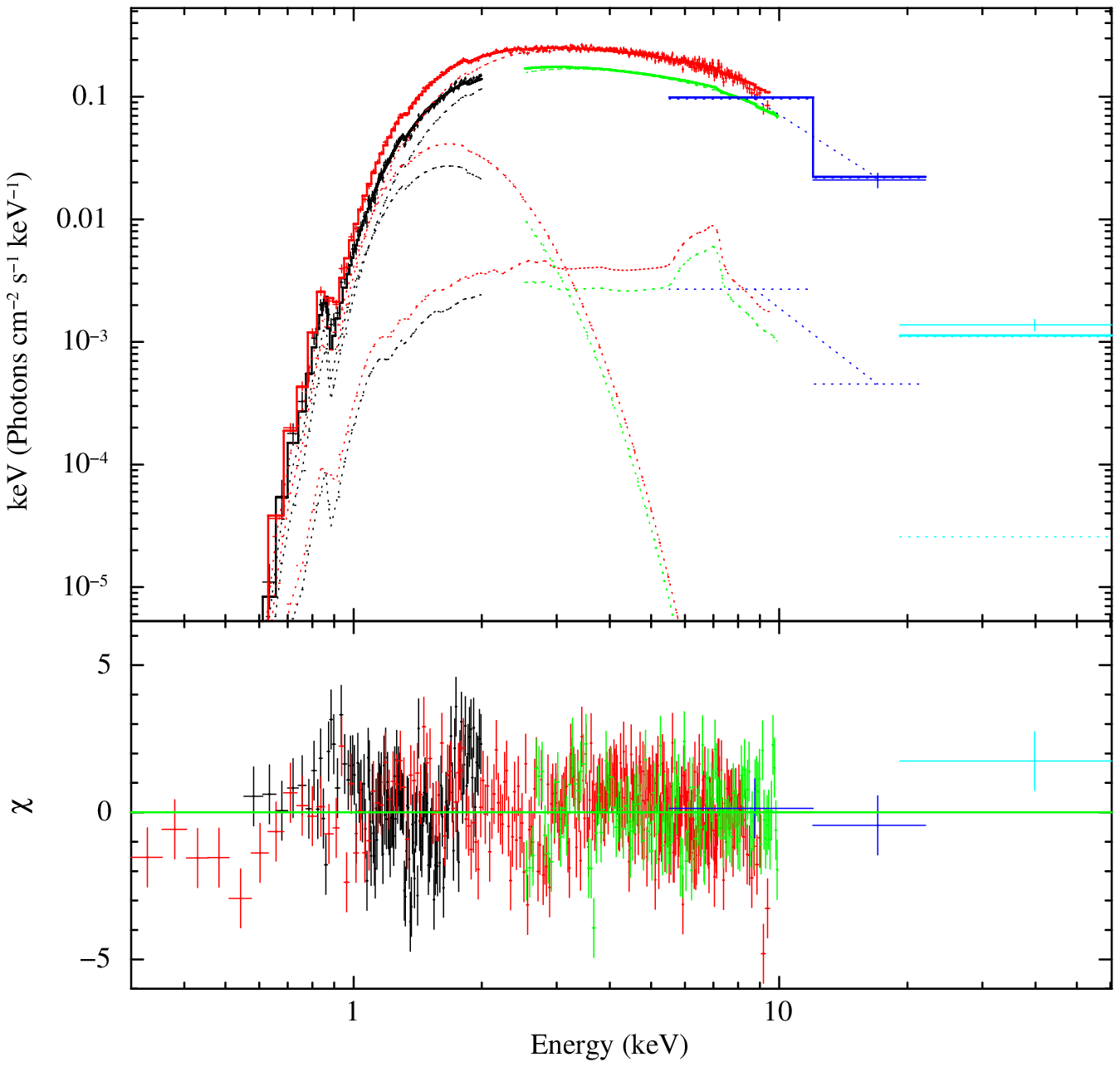}\\
\caption{Left panel: 0.3-60 keV residuals corresponding to the
 trial models  (see Table \ref{fit_results})
  described in the text. Right panel: unfolded spectrum and
  residuals corresponding to the model {\tt CEE+Reflection}. The 
red, black, green, blue, and light-blue data correspond to MOS12,
RGS12, EPIC-pn, JEM-X2, and ISGRI spectra, respectively. The shown data
are graphically rebinned. }
 \label{fit}
\end{figure*} 

\section{Spectral analysis}

Initially, we fitted the data adopting the  model composed of  a
multicoloured disc black-body \citep[{\tt diskbb}  in XSPEC, see][]{Mitsuda_84,Maki_86} plus a
thermal Comptonisation component  \citep[{\tt nthcomp}  in XSPEC,
see][]{zidi_96,zichi_99}.  The parameter {\tt inp\_type} of the {\tt
  nthcomp} 
component was fixed at zero, imposing that the seed-photons of the
Comptonised component describe a black-body spectrum; in our scenario,
the seed-photons  originate from the NS surface  and/or the boundary layer. 
We took into account photoelectric
interstellar absorption using the multiplicative component $\tt phabs,$
assuming the cosmic abundances  and the  photoelectric cross sections
shown by \cite{wilms_00}  and \cite{verner_96}, respectively. 
The fit to the data provided  a $\chi^2(d.o.f.)$ of 2845(2252). The
best-fit values and the corresponding residuals are shown in the third column of Table
\ref{fit_results} and in the left-top
panel of Fig. \ref{fit}. 
Due to the large residuals observed between 0.8 and 1 keV, we added
to the model an absorption edge ({\tt
  edge} in XSPEC) with the threshold energy  and the depth left free to
vary.  We found that the threshold energy associated with
the absorption edge is $0.89 \pm 0.02$ keV with the error at
3$\sigma$ confidence level (c.l.); the obtained $\chi^2(d.o.f.)$ is  2761(2250). The 
 absorption edge  can be only associated with  the \ion{O}{viii} ($\sim$0.871 keV) or
\ion{Ca}{xv} ($\sim$0.895 keV) ions
\citep[see][]{verner_96}. Since \cite{wilms_00} showed that the
abundance of oxygen and calcium are $4.9 \times 10^{-4}$ and $2.19
\times 10^{-6}$, respectively, we conclude that the absorption edge close
to 0.89 keV is associated with the presence of  \ion{O}{viii}
ions. Fixing the threshold energy   at 0.871 keV,
we obtained a $\chi^2(d.o.f.)$ of 2771(2251);
 the improvement from allowing the energy threshold to be free to vary
compared to being fixed gives an
 F-test probability of only  $4.3 \times 10^{-4}$.
 The latter
model (hereafter {\tt model2}) is shown in the fourth column of Table \ref{fit_results}. 
 The addition of the {\tt edge} component to the initial model gave a
 F-test probability of chance improvement of  $1.3 \times 10^{-14}$
 and suggested that its addition  is highly significant.

 The residuals shown in the top-left panel of Fig. \ref{fit} suggest
 the presence of an absorption edge close to 9 keV. Adding this
 component to {\tt model2} (the model composed of the continuum plus
 the two absorption edges is hereafter named {\tt CEE}), we obtained
 a threshold energy of $8.8^{+0.2}_{-0.3}$ keV (3$\sigma$
 confidence level). This value is marginally compatible with the
 energy threshold of an absorption edge associated with
 \ion{Fe}{xxv}  ions.  For this reason, we fitted the spectrum again fixing the
 threshold energy value at 8.829 keV, obtaining a $\chi^2(d.o.f.)$ of
 2729(2250) and a corresponding F-test probability of chance
 improvement of $4.6 \times 10^{-9}$ with respect to {\tt model2}; that is, we achieved
  a statistical improvement at more than  5$\sigma$ c.l..  The best-fit
 parameters and the corresponding residuals are shown in the fifth
 column of Table \ref{fit_results} and in Fig. \ref{fit} (the second
 panel from the top, left side), respectively.  We show the
 corresponding residuals in Fig. \ref{fit} (the second panel from the
 top, left side). There is still a  large excess  in the residuals
 between 6 and 7 keV.

 To fit the residuals between 6 and 7 keV, we added to the {\tt CEE}
 model a Gaussian component.  The addition of the Gaussian line to the
 model substantially changes the best-fit parameters of the continuum
 components.  We found that the inner-disc temperature
 (k$T_{in}= 0.43^{+0.03}_{-0.09}$ keV) is
 compatible with the temperature of the seed photons of the
 Comptonisation (k$T_{bb}=0.61^{+0.09}_{-0.37}$  keV).  Thus, we
 performed a new fit with the k$T_{in} $ and k$T_{bb}$ parameters
 being anchored and imposing the value of the parameter {\tt
   inp\_type} to 1, that is,  assuming that the seed-photons of the
 Comptonising spectrum has a multicoloured disc black-body distribution.
  We obtained a $\chi^2(d.o.f.)$ of 2571(2248) and a
 F-test probability of chance improvement of $5 \times 10^{-30}$ with
 respect to the {\tt CEE} model (see the best-fit values in the sixth
 column of Table \ref{fit_results} and residuals of the fit in
 Fig. \ref{fit}). The Gaussian component has an energy of
 $6.75\pm 0.07$ keV, a width of $0.46 \pm 0.09$ keV, and a
 normalisation of $6.2^{+1.5}_{-1.3} \times 10^{-4}$ photons cm$^{-2}$
 s$^{-1}$; the corresponding equivalent width is $35 \pm 4$ eV.

\begin{table*}
\setlength{\tabcolsep}{3pt} 
\footnotesize
  \caption{Best-fit values of the parameters.\label{fit_results}}
\begin{center}
\begin{tabular}{l l c c c c c c}          
\hline                                             
\hline  
Model &Parameters & Continuum  &Continuum+ &   Continuum+ & CEE+   &       CEE+       &  CEE+ \\  
  &               &            &edge$_1$ &  edge$_1$+edge$_2$ &   Gaussian & diskline & reflection
  \\ 
&                              &      & (model2)  & (CEE) & & & \\
\hline                                             
{\tt PHABS} & N$_H$  ($\times 10^{22}$ cm$^{-2}$) & $2.52 ^{+0.05}_{-0.07}$ &$2.35 \pm 0.05$
& $2.39  \pm 0.05   $ &$2.46 \pm 0.05$  & $2.46 \pm 0.05$ &
                                                                     $2.52
                                                                     \pm
                                                                     0.05$ \\
& & & & \\
{\tt EDGE }& E (keV) & -- & 0.871 (fixed) & 0.871 (fixed) &0.871 (fixed) &0.871 (fixed) &
                                                                       0.871 (fixed) \\
 &$\tau$    & -- &  $0.59 \pm 0.10$&  $0.64 \pm 0.10 $& $0.70 \pm  0.10  $&
            $0.70 \pm 0.10$ &$0.76 ^{+0.09}_{-0.11}$\\ 
& & & & \\
{\tt EDGE }& E (keV) &  -- & --& 8.828 (fixed)& 8.828 (fixed) & 8.828 (fixed) & 8.828 (fixed)\\
                      &$\tau$    & -- &  --& $0.042 \pm    0.010$&  $0.034 \pm 0.011$&$0.038 \pm 0.010$
&$0.037  \pm 0.011$\\
& & & & \\

{\tt DISKBB}&kT$_{in}$   (keV)    & $0.57 \pm 0.04$ & $0.53 \pm 0.04$ &  $0.46 \pm 0.03$
&    $0.37\pm 0.02   $  & $0.37 \pm 0.02$ &$0.34 \pm 0.02 $\\
&R$_{in_{D_{10}}} \sqrt{\cos \theta}$   (km)& $15.1  ^{+2.1}_{-1.4}$ &  $17  \pm 2$  &   $22  ^{+5}_{-3}$
 & $40 \pm 7 $&$40 ^{+8}_{-6}$&$50 ^{+8}_{-10}$
   \\
& & & & \\
{\tt NTHCOMP} &$\Gamma_{comp}$   &  $1.58 \pm 0.02$ & $1.59 \pm
                                                      0.02$  & $1.64
                                                               \pm
                                                               0.02 $
& $1.689   \pm   0.013$ &  $1.692 \pm 0.011 $&$1.710 ^{+0.009}_{-0.014}$\\
& kT$_{bb}$   (keV)      & $<0.26$&$<0.24$&$<0.26$ & $0.37\pm 0.02   $  &  $0.37  \pm 0.02 $
 &$0.34 \pm 0.02 $\\
& kT$_{e}$   (keV)      & $2.27 \pm 0.04$  &$2.30 \pm 0.04$  & $2.44 \pm 0.05$
& $2.55 \pm 0.05$&$2.57 \pm 0.05$ &$2.63 \pm 0.06$
  \\
      &inp$\_$type    & 0 (fixed)& 0 (fixed) & 0 (fixed) & 1 (fixed) & 1 (fixed) & 1 (fixed)  \\ 
& Norm       &  $0.27 \pm 0.03$  & $0.292 ^{+0.012}_{-0.025}$ &  $0.327 ^{+0.012}_{-0.040} $&
$0.35 \pm 0.02 $  &$0.35 \pm 0.02 $&$0.366 ^{+0.011}_{-0.017}$
  \\
& & & & \\

{\tt GAUSSIAN}&E  (keV)   &  -- & --& --&$6.75 \pm 0.07$&--&--\\
&$\sigma$ (keV)   &  -- & --& --&$0.46 \pm 0.09$&--&--\\
&I  ($\times 10^{-4}$ phot. cm$^{-2}$ s$^{-1}$)   &  -- & --& --&$6.2^{+1.5}_{-1.3}$&--&--\\
&   eq. width$_{gauss}$ (eV) &  -- & --& --&$35 \pm 4 $&--&--\\
& & & & \\
{\tt DISKLINE}&E$_{diskline}$ (keV)    &  -- & --&--&--&$6.74 \pm 0.10 $&--  \\
&betor10    &  -- & --&--&--&$-2.9^{+0.3}_{-0.5} $&--  \\
&R$_{in}$   ($R_g$)  &  -- & --&--&--&$24^{+12}_{-8} $&--  \\
&inclination   (degree)  &  -- & --&--&--&$39 \pm 7 $&--  \\
&I ($\times 10^{-4}$ phot. cm$^{-2}$ s$^{-1}$)   &  -- & --&--&--&$6.0^{+1.0}_{-0.5} $&--
  \\
 & eq. width  (eV)  &  -- & --&--&--&$38 \pm 7 $&  --\\
& & & & \\
{\tt RDBLUR}&betor10    &  -- & --&--& --&--&$-3.2 ^{+0.5}_{-5.1}$   \\
&R$_{in} $   ($R_g$)  &  -- & --& --&--&--&$ 31 ^{+25}_{-12}$  \\
&inclination   (degree) &  -- & --&--& --&--&$44 ^{+33}_{-6}$   \\
& & & & \\
{\tt RFXCONV}&Rel$_{refl}$ &  -- & --&--& --&--&$0.072 ^{+0.028}_{-0.013}$ \\
&Log$\xi$   &  -- & --&--& --&--&$2.69 ^{+0.10}_{-0.13}$\\ 
&Fe/solar &  -- & --&--& --&--&2 (fixed)\\
& & & & \\
&$\chi^2$(d.o.f.) &2845(2252)& 2771(2251) &2729(2250)& 2571(2248) &2551(2246)&2572(2246)\\

\hline                                             

\end{tabular}
\end{center} {\small \sc Note} \footnotesize--- The reported errors
are at 90\% confidence level.  
\\
\end{table*}

Motivated by the broadening of the Gaussian component, we investigated
the scenario in which the residuals between 6 and 7 keV are produced
by an emission line smeared by relativistic effects. Thus, we fitted the
spectrum using the  {\tt diskline} component \citep[see][for an accurate description
of the {\tt diskline} component]{fabian_89} instead of the {\tt
  Gaussian} component. We fixed the outer radius of the {\tt diskline}
component at 1000 gravitational radii because of the fit was
insensitive to its change. We started by leaving the inner-disc temperature and
the seed-photons temperature free to vary during the fit. We
obtained k$T_{in}=0.43^{+0.03}_{-0.08}$ keV and
k$T_{bb}=0.60^{+0.10}_{-0.36}$ keV, respectively. Since k$T_{in}$ and
k$T_{bb}$ were compatible within the uncertainties, we fitted the spectrum
with the k$T_{in} $ and k$T_{bb}$ parameters
 being anchored and imposing the value of the parameter {\tt
   inp\_type} to 1.  We found a $\chi^2(d.o.f.)$ of 2551(2246) and
the F-test probability of chance improvement was
$6.1 \times 10^{-32}$ with respect to the {\tt CEE} model  and 
  $1.6 \times 10^{-4}$
  (significance of improvement of $3.1 \sigma$ c.l.)
  with respect to the {\tt CEE+Gaussian} model, respectively.
We show the
corresponding residuals in Fig. \ref{fit} (bottom panel, left
side). The best-fit values are shown in the seventh column of Table
\ref{fit_results}. We found that the energy of the smeared emission
line is $6.74^{+0.08}_{-0.11}$ keV, the power-law dependence of emissivity
(the parameter {\tt betor10} of the {\tt diskline} component) is
$-2.9^{+0.3}_{-0.5}$, the inner radius of the accretion disc at which
the line originates is $24^{+12}_{-8}$ gravitational
radii ($R_g=GM/c^2$), the inclination angle of the binary system is $39^{+6}_{-7}$
degrees and, finally, the equivalent width of the smeared emission line
is $38 \pm 7$ eV.

Finally, we adopted a self-consistent model (hereafter {\tt
  CEE+Reflection}) to take into account the possible presence of a
reflection component from the accretion disc.  The adopted reflection
model {\tt rfxconv} in XSPEC combines the ionised disc table model
\citep[see][]{ross_07} with the Compton reflection code of
\cite{magi_95}. An accurate description of the {\tt rfxconv} model was
shown by \cite{kole_11}. To take into account possible
smearing from relativistic effects, we convolved the {\tt rfxconv} model
with the {\tt rdblur} model. The convolution model {\tt rdblur} has the same
parameters of the {\tt diskline} above defined.  We
tied the seed-photon temperature of the Comptonised component to the
inner-disc temperature, and the outer radius of {\tt
  rdblur} was  fixed at 1000 gravitational radii.
Initially, we left free to vary the value of the abundance of iron, but the
  spectrum does not have  enough statistics to constrain it.  Then,
we fitted the data keeping fixed the value of the parameter at 1, 1.5,
2, 2.5, and 3.  We found 
a slight flattening of
  the residuals close to 7 keV for a value of 2 of the parameter,  while there was not a substantial
change in the residuals for the largest values of the abundance.
 We obtained  a $\chi^2(d.o.f.)$ of 2582(2246) and 
 2572(2246) for a value of   1 and 2 of the parameter,
 respectively.  
 We show the best-fit values for an abundance of iron of 2 in Table \ref{fit_results}. The
spectrum and the corresponding residuals are shown in Fig. \ref{fit}
(right side). We found that the best-fit parameters of {\tt rdblur}
are compatible with those obtained using {\tt diskline}: the ionisation parameter, Log$(\xi)$, of the reflecting
medium is $2.69^{+0.10}_{-0.13}$ and the reflection fraction is $0.072^{+0.028}_{-0.013}$.

\section{Discussion}

We  fitted the broadband spectrum of 4U 1702-429 using three
different models: a) {\tt CEE+Gaussian}, b) {\tt CEE+Diskline,} and c)
{\tt CEE+Reflection}.  Hereafter, the discussion will be focussed on
the latter. In fact, comparing the fitting results reported in Tab. \ref{fit_results} , we observe 
that a {\tt diskline} component  gives a slightly better fit of the iron line profile
with respect to a Gaussian component.

The reflection fraction obtained from the fit,
which is a measure of the solid angle subtended by the reflector as seen
from the Comptonising corona in units of $2 \pi$, is  small with
respect to the values  obtained for these sources \citep[see
e.g.][where values of $\Omega/2\pi$
around $0.2-0.3$ are found]{disalvo_15, pintore_15} , but it is compatible
with the  values shown by \cite{cackett_2010} for Ser X-1 and 4U
1820-30; moreover, a similar value of the reflection fraction  ($\Omega/2\pi =0.094
\pm 0.008$)
 has been measured in the case of the accreting
millisecond pulsar SAX J1748.9-2021 \citep[see][]{pintore_16}.
Such a small reflection amplitude might indicate a small solid angle
subtended by the reflector as seen from the corona, caused, for instance,
by a quite compact corona with no superposition with the disc, or a
patchy corona above the inner part of the disc, or by a (mildly)
relativistic velocity of the corona away from the disc.

The self-consistent model should include also the absorption edges
associated with transitions of ionised iron;  however, we needed to add an absorption edge
associated with highly ionised iron (transition of \ion{Fe}{xxv}). 
This incongruity was discussed by \cite{Egron_13} studying the broadband
 spectrum of 4U 1705-44 in the soft state. 
In that case, the authors fitted the broadband spectrum of the source using the 
self-consistent models  {\tt reflionx} and 
{\tt xillver}   reproducing the spectrum; nevertheless, the authors needed to add
an absorption edge at $8.7 \pm 0.1$ keV and $8.5 \pm 0.1$ keV (using  the  {\tt
   reflionx}  and {\tt xillver} model, respectively) and suggested that the
 high energy resolution and the large statistics available 
 with the
 XMM-Newton observatory require an improvement of the self-consistent
 model to well account for the threshold energy of the absorption edges
 associated with ionised iron.

\cite{galloway_08}, assuming a neutron star mass of 1.4 M$_{\odot}$,
estimated as a possible distance to the source  $4.19 \pm 0.15$ and
$5.46 \pm 0.19$ kpc for pure hydrogen and pure helium  companion stars,
respectively.   Furthermore, the same authors suggested that the
companion star should have a mass fraction of hydrogen, $X$, of less than
50\% (i.e. $X<0.35$) because the
observed bursts have short rise times ($<2$ s) and low $\tau$-values
($\tau=7.8 \pm 0.7$ s).  Using this argumentation, 
the lower limit of the distance is $4.70 \pm 0.16$ kpc 
for   $X=0.35$.

We  estimate the distance
to the source using the obtained value of the equivalent hydrogen
column, $N_H$ of the interstellar matter.  \cite{guver_09} show that
$N_H$ is related to the visual extinction $A_V$ from the
relation
$$
N_H=(2.21 \pm 0.09) \times 10^{21} A_V {\rm \;\;cm^{-2}},
$$
while $A_V$ is related to the infrared extinction in the $K_s$ band
through the relation
$$
A_{K_{s}}=(0.062 \pm 0.005) A_V {\rm \; mag,}
$$
\citep[see][]{nishi_08}. Combining the last two relations we find  
$N_H = (3.6 \pm 0.3) \times 10^{22} A_{K_{s}}$.  Using the
value of $N_H$ obtained by our fit  and combining these two relations,
we find
that $A_{K_{s}}= 0.70 \pm 0.06$.
We take into account the 3D extinction map
of the radiation in the K$_s$ band for our Galaxy
\citep{marshall_06} to infer the distance to the
source. We use the available radial profile of A$_{k_{s}}$  
closest to 4U 1702-429 obtained for  
$l=344^{\circ}$ and $b=-1.25^{\circ}$ (the Galactic coordinates of 4U 1702-429 are
$l=343.9^{\circ}$ and $b=-1.32^{\circ}$). 

\begin{figure}
\centering
\includegraphics[width=8.5cm]{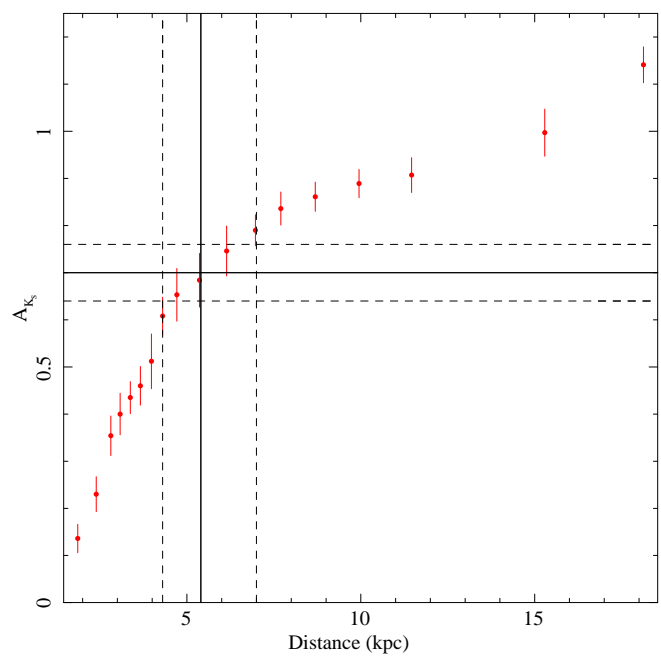}\\
\caption{Infrared extinction  $A_{K_{s}}$ versus the distance
  along the  Galactic
  coordinates  $l = 344^{\circ}$ and $b=-1.25^{\circ}$ \citep{marshall_06}.
The solid and dashed horizontal lines indicate the values of   $A_{K_{s}}$ and
the corresponding errors that we infer from the equivalent hydrogen column density $N_H$ obtained from the fit. }
 \label{aks}
\end{figure}

We show A$_{k_{s}}$ versus
the distance in Fig. \ref{aks}.
In the same figure, the  horizontal solid line indicates the best-value of
A$_{k_{s}}$
obtained above and the   horizontal dashed lines indicate the values of A$_{k_{s}}$
taking into account the associated error at 90\% of confidence level. 
The vertical lines indicate the corresponding range of the distance to
the source.
We find a distance to the source of $5.4^{+1.6}_{-1.1}$ kpc that 
is compatible with the range   4.5-5.7 kpc 
suggested by \cite{galloway_08} and indicates  that our estimate of the
equivalent hydrogen column density of the interstellar medium is
correct.  The value of  $N_H \sim  1.7 \times 10^{22}$ cm$^{-2}$
shown by \cite{Cristian_97}  implies a distance to the source smaller
than 4.3 kpc, while a value of  $N_H \sim  0.9 \times 10^{22}$
cm$^{-2}$  suggested by \cite{Oo_91} implies a distance  
smaller than 2.8 kpc. We adopt hereafter a distance to the source of 5.5
kpc.

We extrapolate an unabsorbed flux of $4 \times 10^{-9}$  erg s$^{-1}$ cm$^{-2}$ in the 0.1-100
keV energy band, corresponding to an unabsorbed luminosity of
$1.5 \times 10^{37}$ erg s$^{-1}$.  For comparison with
\cite{mark_99}, we extrapolate the unabsorbed flux in the 1.6-30 keV
($1.6\times 10^{-9}$ erg s$^{-1}$ cm$^{-2}$) which is in line with
their findings.  Finally, we extrapolate the unabsorbed flux  of the
Comptonisation spectrum, $f_{bol}$, obtaining 
$3.4 \times 10^{-9}$ erg s$^{-1}$ cm$^{-2}$ in the 0.1-100 keV energy band.

The $R_{in}$ value shown in Table
\ref{fit_results}
represents  the apparent  inner  radius of the accretion disc.  
We apply  a correction factor 
to convert the apparent inner  radius $R_{in}$  into the realistic inner
radius  $r_{in}$ \citep[see][for a detailed discussion]{kubota_98}. The relation between
$r_{in}$ and $R_{in}$ is $r_{in}=\alpha \kappa^2 R_{in}$, where 
$\kappa \simeq 1.7$ for a luminosity close to  10\% of the
Eddington luminosity \citep[see ][]{shimu_95} and $\alpha \simeq 0.41$
for a zero-torque condition  \citep{kubota_98}. We obtain  $r_{in}=
39^{+6}_{-8}$ km using the inclination
  angle of $44^{\circ}\!$ obtained from the best-fit and adopting a
  distance to the source of 5.5 kpc. Furthermore, the inner radius,
  $R_{refl}$,  where the reflection component originates, is $ 31 ^{+25}_{-12}$ 
gravitation radii (see Table \ref{fit_results}) that,  assuming a NS
 mass of   1.4 M$_\odot$, corresponds to a radius of   $ 64 ^{+52}_{-15}$  km.

To estimate the optical depth $\tau$ of the Comptonising region, we
adopted the relation
$$
\Gamma_{comp} = \left[ \frac{9}{4}+ \frac{1}{\tau (1+\frac{\tau}{3})\left(\frac{kT_e}{m_ec^2}\right)} \right]^{1/2}-\frac{1}{2}
$$ 
\citep[see][]{zidi_96}, where $m_e$ and $c$ are the  
electron rest mass and the speed of light, respectively. Using the
best-fit values of $\Gamma_{comp}$ and k$T_e$ we obtain
$\tau = 13.6 \pm 0.2$.  We estimate the  emission radius of the seed
photons, assuming
a spherical emission and that most
of the seed photons are scattered in the optically thick corona. Following
\cite{zand_99}, the emission radius of the seed
photons in units of km can be calculated as $R_0 = 3 \times 10^4
d \;[f_{bol}/(1+y)]^{1/2} (kT_0)^{-2}$, where $d$ is the source distance
in units of kpc, $y$, defined as $y=4kT_e\tau^2/(m_ec^2)$, is the Compton parameter and
$kT_0$ is the seed photons temperature; in our case $kT_0=kT_{in}$. 
We find that $R_0 =38 \pm 7$ km, assuming a conservative error of 10\%
associated with  $f_{bol}$. We conclude that the values of 
$R_{in}$,  $R_0$ , and $R_{refl}$ are consistent with each other.

Assuming that the Comptonising region has a
spherical geometry surrounding the NS and that it has a roughly
homogeneous density, we estimate the electron density $n_e$ from
$\tau=n_e \sigma_T l$, where  $\sigma_T$  and $l$ are the
Thomson cross-section and the geometrical size of the
Comptonising plasma, respectively. Assuming that the region extends from the NS
surface up to the observed inner accretion disc radius, we find that
$n_e = 5.2^{+0.9}_{-1.2} \times 10^{18}$ cm$^{-3}$.

We also estimate the electron density $n_e$ of the reflecting skin
above the accretion disc using the relation $\xi = L_x/(n_er^2)$,
where $L_x$ is the unabsorbed incident luminosity in the 0.1-100 keV
energy range, $\xi$ is the ionisation parameter (see Table
\ref{fit_results}), and $r$ is the inner radius of the disc
where the reflection component originates, $R_{refl}$.  Since we
find  that $L_x \simeq 1 \times 10^{37}$ erg s$^{-1}$, we obtain  $n_e \sim 5 \times 10^{20}$ cm$^{-3}$  adopting $r=64$
km. The electron density associated with the reflecting skin is a
factor of one hundred larger than that associated with  the
Comptonised corona surrounding the NS.
 Because the reflecting region above the disc has to be optically thin, using
the relation $n_e \sigma_T l <1$,  we  find that the upper limit to
its geometrical depth is 0.03 km. 

The height, $H$, of the accretion disc is 
$H=1.7 \times 10^8
\alpha^{-0.1}\dot{m}_{16}^{3/20}m_1^{-3/8}r_{10}^{9/8}$ cm \citep[see
eq. 5.46 in ][]{frank_2002}, where 
$\dot{m}_{16}$ is the mass accretion rate in units of $10^{16}$ g/s, 
$m_1$ is the NS mass in units of solar masses, and 
$r_{10}$ is the radius of the accretion disc in units of $10^{10}$ cm.
Adopting  $\alpha=0.1$ and using   $\dot{m}_{16} \simeq 8$ g/s 
(obtained from the
unabsorbed flux showed above), we find that $H \simeq 0.65$ km for a radius of
64 km. This implies that the height of the accretion disc is 
a factor of 20 larger than the reflecting skin at a radius of 64 km.

We detect two absorption edges 
associated with \ion{O}{viii} and \ion{Fe}{xxv} ions, respectively. 
We estimate that the corresponding photoionisation cross-section
values are $\sigma_{O8}= 9.9 \times 10^{-20}$ and     $\sigma_{Fe25}=2.2 \times 10^{-20}$
cm$^{2}$ \citep[see][]{verner_96}. 
The observed optical depth $\tau$
associated with the absorption edge of the   \ion{Fe}{xxv} is
$\tau=\sigma_{Fe25}n_{Fe25}l$, where   $n_{Fe25}$ is the number
density of
\ion{Fe}{xxv} ions  and $l$ is the geometrical size along the line of
sight of the
region where the absorption edge originates. 
The corresponding equivalent column of \ion{Fe}{xxv} can be written as
$N_{Fe25}=n_{Fe25}l$ and the optical depth can be re-written as
$$
\tau= \sigma_{Fe25} \frac{N_{Fe25}}{N_{Fe}}\frac{N_{Fe}}{N_H} N_H,
$$    
where $f_{Fe25}=N_{Fe25}/N_{Fe}$ is the fraction of  \ion{Fe}{xxv} ions
with respect to whole population of iron-ions  
and $N_{Fe}/N_{H}$ is the
abundance of iron with respect to hydrogen abundance. Adopting the
cosmic abundances shown by 
 \cite{wilms_00} 
$N_{Fe}/N_{H}$  is $2.69 \times 10^{-5}$. 
 The parameter $f_{Fe25}$ 
is inferred using the results of  \cite{kallman_01} for an electron
density  equal or larger than $10^{17}$ cm$^{-3}$ and an ionisation
parameter of Log$(\xi)=2.69$ as obtained from the fit. We find that  
$f_{Fe25} \simeq 0.50$.  Considering that the spectral fit indicates
an iron abundance twice that of the cosmic one, we
find that  the corresponding equivalent hydrogen column
associated with the \ion{Fe}{xxv} ion is $N_H=(6 \pm 2) \times
10^{22}$ cm$^{-2}$. 

Assuming that the absorption edge associated with the \ion{O}{viii}
ion originates from a region in the accretion disc surface  close
to the region where the  smeared relativistic line and the absorption
edge of \ion{Fe}{xxv} originate,  we can roughly assume that the
equivalent column of neutral hydrogen associated with the  \ion{O}{viii}
ion  is the same as that inferred for the  \ion{Fe}{xxv} ion. 
Using the relation 
$$
\tau= \sigma_{O8} \frac{N_{O8}}{N_{O}}\frac{N_{O}}{N_H} N_H,
$$
we  estimate $f_{O8}=N_{O8}/N_{O}$ knowing that 
$N_{O}/N_H$ is $4.9 \times 10^{-4}$ \citep[see][]{wilms_00}.
We obtain $f_{O8} \simeq 0.26$ that,  for an electron
density  equal or larger than $10^{17}$ cm$^{-3}$, corresponds to an
ionisation parameter of Log$(\xi) \simeq 1.9$
\citep[see][]{kallman_01}.

Even if the absorption edges associated with the \ion{O}{viii} and
\ion{Fe}{xxv} ions originate in close regions above the accretion disc,
we should observe only the smeared relativistic iron line associated with the
\ion{Fe}{xxv} ions because the
fluorescence yield $F$ depends on $Z^4/(30^4+Z^4)$, where $Z$ is the
atomic number. $F$ is $5 \times 10^{-3}$ and 0.36 for oxygen and iron,
respectively; weighting the fluorescence yield for the cosmic
abundances and the fraction $f$ of \ion{O}{viii} and \ion{Fe}{xxv}
 shown above, we obtain that $F$ is $6.4 \times 10^{-7}$ and
$1 \times 10^{-5}$ for oxygen and iron, respectively. This means that
the strength of the smeared relativistic line associated with the
\ion{O}{viii} ion should be only 6.6\% of that associated with the
\ion{Fe}{xxv} ion. The large interstellar photoelectric absorption
does not allow us to significantly detect the presence of the smeared
relativistic line associated with the \ion{O}{viii} ion. To test this,
we  added an
additional {\tt diskline} model to the {\tt CEE+Diskline} model, fixing
the smearing parameters to that of the smeared relativistic iron line. This does not include
the energy which has been fixed at 0.6536 keV, which is the expected
energy for an emission line associated with  \ion{O}{viii}.  We find an upper limit
to the normalisation of 0.17  phot. cm$^{-2}$ s$^{-1}$  at 3$\sigma$
c.l., and the fit is not improved by adding
this further component.

\section{Conclusions}

In this work, we have shown the first broadband spectral analysis of the persistent spectrum of \1702
in the 0.3-60 keV energy range. We detect the presence of a prominent
feature close to 6.7 keV and two absorption edges at 0.87 and 8.83
keV, respectively. The emission line at 6.7 keV is associated with the
fluorescence emission of \ion{Fe}{xxv} ions and it can be modelled
using a Gaussian component with $\sigma=0.46$ keV and an equivalent
width of 36 eV.  Alternatively, it can be described by a relativistic
smeared line caused by Compton reflection originating from the
inner disc.  We have fitted the spectrum with a self-consistent model
composed of a multicoloured disc black-body component plus a
Comptonisation component and a reflection component.  We find that the
inclination angle of the system is $44^{\circ}\!$, the inner
radius of the accretion disc is $39^{+6}_{-8}$ km, and its inner
temperature is 0.34 keV.  The inner radius of the reflecting region,
where the \ion{Fe}{xxv}  smeared relativistic line and the
\ion{Fe}{xxv}  absorption edge are produced, is $64^{+52}_{-15}$ km, and
the corresponding ionisation parameter is Log$(\xi)=2.69$.  The
ionisation parameter of the reflecting region where the absorption
edge associated with \ion{O}{viii} ions originates is  Log$(\xi) \sim
1.9$.  The electron temperature of the
Comptonised component is 2.6 keV and the corresponding optical depth
is $\tau \sim 13.6$.
 
From the best-fit value of the equivalent hydrogen column density of the
interstellar medium ($N_H \sim 2.5 \times 10^{22}$ cm$^{-2}$), we have 
estimated the infrared extinction $A_{K_{s}} = 0.70 \pm 0.06$ mag and
inferred a distance to the source of $5.4^{+1.6}_{-1.1}$ kpc. This value
is compatible with a previous estimation obtained from the analysis of the
photospheric radius expansion of the type-I X-ray bursts observed for
\1702.

\section*{Acknowledgments}
This research has made use of data and/or software provided by the
High Energy Astrophysics Science Archive Research Center (HEASARC),
which is a service of the Astrophysics Science Division at NASA/GSFC
and the High Energy Astrophysics Division of the Smithsonian
Astrophysical Observatory.  This research has made use of the VizieR
catalogue access tool, CDS, Strasbourg, France.  The High-Energy
Astrophysics Group of Palermo acknowledges support from the Fondo
Finalizzato alla Ricerca (FFR) 2012/13, project N. 2012-ATE-0390. We also acknowledge a financial
contribution from the agreement ASI-INAF I/037/12/0.  MDS thanks the
Dipartimento di Fisica e Chimica, Università di Palermo, for its
hospitality.  AR and AS gratefully acknowledge the Sardinia Regional
Government for  its financial support (P.O.R. Sardegna
F.S.E. Operational Programme of the Autonomous Region of Sardinia,
European Social Fund 2007-2013 - Axis IV Human Resources, Objective
l.3, Line of Activity l.3.1.). AP acknowledges grant SGR2014-1073 and
partial support from “NewCompStar”, COST Action MP1304.
\bibliographystyle{aa} 
\bibliography{citations}
\end{document}